\documentclass[aps,prd,twocolumn,preprintnumbers,superscriptaddress,nofootinbib]{revtex4-1}
\usepackage{graphicx}
\usepackage{epstopdf}
\usepackage{amsmath}
\usepackage{amsfonts}
\usepackage{amssymb}
\usepackage{appendix}
\usepackage{enumerate}
\usepackage{natbib}
\usepackage{comment}
\usepackage{bbold}
\usepackage[shortlabels]{enumitem}
\usepackage{color}
\usepackage{slashed}
\usepackage{subfigure}
\usepackage{setspace}
\usepackage{footnote}
\usepackage{lipsum}
\usepackage{multirow}
\usepackage[colorlinks = true,
            linkcolor = blue,
            urlcolor  = blue,
            citecolor = blue,
            anchorcolor = blue]{hyperref}
\usepackage[capitalize]{cleveref}
\usepackage{braket}
\usepackage[compat=1.1.0]{tikz-feynman}
\usepackage{multirow}
\usepackage{physics}
\usepackage{feynmp-auto}
\usepackage[normalem]{ulem}
\usepackage{url}
\usepackage{units}
\usepackage[normalem]{ulem}

\newcommand{\be}{\begin{eqnarray}}
\newcommand{\ee}{\end{eqnarray}}
\newcommand{\ba} {\begin{equation}\begin{aligned}}
\newcommand{\ea} {\end{aligned}\end{equation}}
\newcommand{\bg} {\begin{equation}\begin{gathered}}
\newcommand{\eg} {\end{gathered}\end{equation}}

\newcommand{\beq}{\begin{equation}}
\newcommand{\eeq}{\end{equation}}

\usepackage{tikz,xcolor,hyperref}

\definecolor{lime}{HTML}{A6CE39}
\DeclareRobustCommand{\orcidicon}{\hspace{-1mm}
	\begin{tikzpicture}
		\draw[lime, fill=lime] (0,0) 
		circle [radius=0.12] 
		node[white] {{\fontfamily{qag}\selectfont \tiny \,ID}};
		\draw[white, fill=white] (-0.0525,0.095) 
		circle [radius=0.007];
	\end{tikzpicture}
	\hspace{-3mm}
}

\foreach \x in {A, ..., Z}{\expandafter\xdef\csname orcid\x\endcsname{\noexpand\href{https://orcid.org/\csname orcidauthor\x\endcsname}
		{\noexpand\orcidicon}}
}

\begin{document}

\title{Prospects for Neutrino Observation and Mass Measurement\\ from Binary Neutron Star Mergers}

	\author{Vedran~Brdar\orcidA}
	\email{vedran.brdar@okstate.edu}
	\affiliation{Department of Physics, Oklahoma State University, Stillwater, OK 74078, USA}
	\author{Dibya~S.~Chattopadhyay\orcidB}
	\email{dibya.chattopadhyay@okstate.edu}
	\affiliation{Department of Physics, Oklahoma State University, Stillwater, OK 74078, USA}
    \author{\mbox{Samiur~R.~Mir\orcidC}}
	\email{samiur.mir@okstate.edu}
	\affiliation{Department of Physics, Oklahoma State University, Stillwater, OK 74078, USA}
    \author{Tousif~Raza\orcidD}
	\email{tousif@okstate.edu}
	\affiliation{Department of Physics, Oklahoma State University, Stillwater, OK 74078, USA}
    \author{Marc~S.~Romanowski\orcidE}
	\email{marc.s.romanowski@gmail.com}
	\affiliation{The College of New Jersey, Ewing Township, NJ 08618, USA}
	\affiliation{Department of Physics, Oklahoma State University, Stillwater, OK 74078, USA}

\begin{abstract}
Over the next decade, $\mathcal{O}(100)$ diffuse supernova neutrino background (DSNB) events are expected to be detected in the upcoming Hyper-Kamiokande experiment. Another neutrino source that has received far less attention in the context of neutrino detection is binary neutron star mergers. Including the data from multiple recent simulations, we find that detection in current and near-future neutrino experiments is not feasible, and a megaton-scale detector with $\mathcal{O}(10)$ MeV threshold, such as the proposed Deep-TITAND, MEMPHYS, or MICA, will be required. This is primarily due to the updated binary neutron star merger rate and the time-of-flight delay caused by the nonzero neutrino mass. Regarding the former, recent results from the fourth observing run by LIGO, Virgo, and KAGRA has significantly lowered the upper limit on the neutron star merger rate. As for the latter, neutrino events from neutron star mergers are expected to be recorded shortly after the gravitational wave signal associated with the coalescence. Limiting the analysis to such short time windows can significantly reduce background rates. While this approach has been qualitatively discussed in the literature, the effect of the time delay caused by nonzero neutrino mass, which can substantially extend the observation windows, has been disregarded. We present a refined analysis employing energy-dependent time windows and luminosity distance cuts for the mergers and provide realistic estimates of the detector runtime required to record neutrinos from binary neutron star mergers with small background contamination. The relative timing between the neutrino and gravitational wave signals can also be employed to probe the scale of neutrino mass. We find that the sensitivity to the lightest neutrino mass exceeds both the most stringent terrestrial bounds from KATRIN and the projections based on galactic supernovae. This level of sensitivity may become particularly relevant in the future if terrestrial and supernova constraints are not significantly improved.
\end{abstract}

\maketitle
\section{Introduction}
\label{sec:intro}
\noindent
The discovery of neutrino oscillations \cite{Super-Kamiokande:1998kpq, SNO:2002tuh} is indisputable evidence for the massive nature of neutrinos. While this type of experiment is only sensitive to neutrino mass squared differences, there are complementary experimental techniques for probing neutrino mass. Namely, the KATRIN experiment \cite{KATRIN:2021dfa} measures the spectrum of electrons from molecular tritium decay, from which the effective electron antineutrino mass, \mbox{$m_{\nu_e} \equiv \left(\sum_i |U_{ei}|^2 m_i^2\right)^{1/2}$}, is currently constrained to $m_{\nu_e}<0.45~\text{eV}$ (90\% CL) \cite{KATRIN:2024cdt}. Here, $m_i$ denotes the neutrino masses and $U_{ei}$ are elements of the first row of the leptonic mixing matrix, also known as the Pontecorvo-Maki-Nakagawa-Sakata (PMNS) matrix. This limit translates to a direct limit on neutrino mass. In neutrinoless double beta decay, another quantity involving PMNS elements and neutrino masses is probed for Majorana neutrinos: the effective neutrino mass $|m_{ee}| \equiv \left|\sum_i U_{ei}^2 m_i\right|$. At present, the most stringent limit on this parameter is set by KamLAND-Zen \cite{Shirai:2017jyz} and is of order $0.1~\text{eV}$ \cite{KamLAND-Zen:2024eml}, with the precise value being hindered by uncertainties in nuclear matrix element calculations. It should be noted, however, that this constraint does not apply if neutrinos are Dirac particles.
Yet another probe sensitive to the neutrino mass comes from cosmological observations, which can constrain the sum of neutrino masses. Constraints from Planck give $\sum_i m_i < 0.12~\text{eV}$ (95\% CL) \cite{Planck:2018vyg}, while more recent results from DESI \cite{DESI:2016igz} yield $\sum_i m_i < 0.053~\text{eV}$ (95\% CL) \cite{DESI:2025ejh} within the $\Lambda$CDM model. The latter result, even in the normal neutrino mass ordering, and for the lightest neutrino mass set to $0~\text{eV}$, is in tension with terrestrial measurements of neutrino mass obtained from oscillation experiments. It should be noted that the cosmological constraints on neutrino mass can get significantly relaxed in models beyond $\Lambda$CDM and in scenarios where neutrinos have non-standard interactions \cite{Beacom:2004yd,Bellomo:2016xhl,Esteban:2021ozz,Barenboim:2020vrr,Escudero:2020ped}, highlighting strong model dependence.

Neutrino mass can also be constrained by analyzing arrival times and energies of supernova neutrinos by leveraging neutrino time-of-flight effects \cite{Zatsepin:1968kt}. This was done for SN1987A in Refs.~\cite{Loredo:2001rx,Pagliaroli:2010ik}, where an upper bound on the neutrino mass of $5.8$ eV (95\% CL) was derived. Further improvements on this limit are expected with neutrino data from the next galactic supernova \cite{Totani:1998nf,Beacom:2000ng,Nardi:2003pr,Pompa:2022cxc,Parker:2023cos}. A galactic supernova may also be detected in gravitational waves \cite{Szczepanczyk:2021bka}. In that case, neutrino mass can also be constrained by using the difference in detection times of gravitational waves and neutrinos \cite{Arnaud:2001gt,Hansen:2019giq}, since the latter travel at speeds smaller than the speed of light given their nonzero mass; see also \cite{Nishizawa:2014zna,Langaeble:2016han} for a general framework for timing neutrino and gravitational wave signals. While the time delay of the neutrino signal grows for larger neutrino mass, it also scales linearly with the baseline. This suggests that neutrino sources beyond our galaxy, which offer simultaneous detection through gravitational waves, would be highly sensitive to neutrino mass. In this work, we show that one of the primary candidates for such sources are binary neutron star mergers, which occur at much larger distances compared to the $\sim 10$ kpc distance from galactic supernova candidates; note that the two neutron star merger events detected by LIGO \cite{LIGOScientific:2017vwq,LIGOScientific:2020aai} occurred at distances of $\mathcal{O}(100)$ Mpc from Earth.

Using neutron star mergers to constrain the neutrino mass was discussed for the first time in Ref.~\cite{Kyutoku:2017wnb}, where the authors found that Hyper-Kamiokande~\cite{Hyper-Kamiokande:2018ofw} would need to take data for over 100 years to record a single neutrino event from these sources. The authors pointed out that background events can be strongly reduced by considering only $\mathcal{O}(1)$ s time windows following each neutron star merger observed through gravitational waves. The idea is that the majority of neutrinos from binary neutron star mergers are emitted within such a short time interval, and therefore reducing the total time considered in a given analysis significantly lowers the background while keeping the signal (neutrino event count from binary neutron star mergers) practically unaltered. The authors of Ref.~\cite{Lin:2019piz} calculated the expected neutrino event rates using data from several available simulations and also increased the number of considered binary neutron star mergers by including higher redshifts in their analysis. While the 200 Mpc cutoff considered in Ref.~\cite{Kyutoku:2017wnb} is around the maximum distance at which present-day ground-based gravitational wave detectors can observe binary neutron star mergers, Cosmic Explorer \cite{Reitze:2019iox} and the Einstein Telescope \cite{Abac:2025saz} will indeed be sensitive to significantly larger distances. This allowed the authors of Ref.~\cite{Lin:2019piz} (see also \cite{Deguire:2025zmg}) to obtain more optimistic predictions for the neutrino event rates compared to Ref.~\cite{Kyutoku:2017wnb}; they also compared signal and background events to infer the capabilities of upcoming neutrino detectors to identify such events.

In our work, we complement and extend the analyses presented in Refs.~\cite{Kyutoku:2017wnb,Lin:2019piz,Deguire:2025zmg}. First, we update the predicted neutrino event rates from neutron star mergers by considering results from recent simulations featuring neutrino luminosities; see the review \cite{Foucart:2022bth}. We also perform calculations using the new upper limits on the local neutron star merger rate from the first part of the fourth observing run of LIGO, Virgo \cite{VIRGO:2014yos}, and KAGRA \cite{KAGRA:2020tym} (LVK), reported to be $250\, \text{Gpc}^{-3}\,\text{yr}^{-1}$ in Ref.~\cite{LIGOScientific:2025pvj}. This significant reduction of the upper limit compared to $1700\, \text{Gpc}^{-3}\,\text{yr}^{-1}$, corresponding to the first three LVK runs \cite{KAGRA:2021duu}, is crucial for the detection prospects at neutrino experiments. Namely, we will show that Hyper-Kamiokande will not be able to achieve a successful detection of neutrinos from binary neutron star mergers within a reasonable operational time frame, and therefore, larger, megaton-scale detector(s) will be necessary.

Second, we include the effect of neutrino mass in our background mitigation analysis. Namely, while the vast majority of neutrinos from binary neutron star mergers may indeed be emitted within $\mathcal{O}(1)~\textrm{s}$ after the coalescence, due to nonzero neutrino mass, the neutrino time delay at Earth, $\Delta t \approx L\, m_\nu^2/(2 E_\nu^2)$, for a typical neutrino energy of $E_\nu = 20$ MeV, a merger distance of $L = 300$ Mpc, and a neutrino mass at the current KATRIN limit of $m_\nu = 0.45$ eV, is approximately $\Delta t \approx 7$ s. Such a time delay is about an order of magnitude larger compared to the typical neutrino emission time period from simulations, and this already demonstrates that the signal-to-background ratio in \cite{Kyutoku:2017wnb,Lin:2019piz,Deguire:2025zmg} is overestimated.

Third, we perform a detailed analysis to infer the projected sensitivity on neutrino mass using neutrinos from binary neutron star mergers. The neutrino mass constraint that can be obtained would depend on the luminosity distance, neutrino energy, and the observed time delay. In this analysis, which features all relevant uncertainties, such as those associated with the neutrino emission time and distances to the mergers, we find that even $\mathcal{O}(0.1)$~eV neutrino mass can be constrained. This result is an order of magnitude weaker compared to the simple estimates presented in \cite{Kyutoku:2017wnb}, where $\mathcal{O}(0.01)$ eV capability was discussed.

This work is organized as follows.
\cref{sec:events,sec:bkg,sec:mass} are dedicated to each of the three above components, namely, neutrino event rate, background mitigation analysis, and neutrino mass sensitivity, respectively. In \cref{sec:conclusions}, we conclude.

\section{Neutrino event rate}
\label{sec:events}
\noindent
Following a neutron star merger, neutrinos can be emitted either through the cooling of the newly formed neutron star or from the accretion disk \cite{Foucart:2022bth}. In over 60\% of binary neutron star mergers, a neutron star that survives for 1~s or longer is formed \cite{Sarin:2020gxb,Gao:2015xle,Li:2024luu,Margalit:2019dpi,Beniamini:2021tpy,Piro:2017zec}. Even in cases where a black hole is formed promptly after the merger, an accretion disk still develops and emits neutrinos (see, for instance, Ref.~\cite{Schilbach:2018bsg}). In this work, we do not include the latter contribution and consider only neutrinos from the post-merger neutron star and its corresponding accretion disk. The neutrino luminosities in these cases are expected to be larger than in black hole scenarios\footnote{\color{black}{The prompt black hole formation occurs in $\sim 35$\% of all mergers~\cite{Li:2024luu}}.} by at least a factor of a few; see for instance the total neutrino luminosity (neutron star cooling plus accretion disk contribution) in Fig.~3 of \cite{Fujibayashi:2020dvr} and compare it with the accretion disk contribution in the black hole scenario shown in Fig.~1 of \cite{Fujibayashi:2020jfr}. Another example of that is given in \cite{Lippuner:2017bfm}, where the ``Hinf'' model features an overall dominance in neutrino luminosity compared to models that lead to earlier collapse into a black hole. While the black hole contribution could still slightly enhance our signal prediction, the main reason for not considering it is related to background mitigation.
Neutron star mergers within a distance of at most a few hundred Mpc that result in a promptly formed black hole can be distinguished from those producing a neutron star that survives for 1 s or longer, through the post-merger gravitational wave signal{\color{black}, observable in the future gravitational wave detectors} \cite{Srivastava:2022slt,Weih:2019xvw,Topolski:2023ojc,Grace:2024xty,Lehoucq:2025ruc}, as well as through a possible photon signal \cite{LIGOScientific:2017ync} expected in the latter case. Tagging only the subset of mergers that do not result in a promptly formed black hole implies that the average time between mergers of interest will be longer than if all mergers were considered. This results in a further mitigation of the background, without a significant reduction of the neutrino flux from the mergers; see details in \cref{sec:bkg}.

Simulations of binary neutron star mergers yield neutrino luminosities and average neutrino energies, which are key ingredients for computing the neutrino fluxes. In Ref.~\cite{Lin:2019piz}, the authors extracted these quantities from several simulations where the merger results in either a hypermassive neutron star or a prompt collapse into a black hole. Regarding the former, and focusing on the electron antineutrino luminosity, the simulations in Refs.~\cite{Lippuner:2017bfm,Fujibayashi:2017xsz,Sekiguchi:2011zd} report results that span an order of magnitude in this quantity (see also table in Ref.~\cite{Lin:2019piz}), which in turn implies an order-of-magnitude uncertainty in the neutrino flux.
In more recent simulations, the total neutrino luminosity is shown for $6$ different models in Fig.~13 of Ref.~\cite{Shibata:2021xmo} and compared to the results of a viscous-hydrodynamics simulation in Ref.~\cite{Fujibayashi:2020dvr}, where the latter also featured in a review~\cite{Foucart:2022bth}. The results are comparable; across the provided time interval, the neutrino luminosities from these different simulations differ by at most a factor of $3$. We also note results from another group~\cite{Radice:2021jtw,Qiu:2025kgy}, where luminosities beyond $\sim 0.1$ s after the merger are not presented. In Ref.~\cite{Lin:2019piz}, for several models, the authors addressed this by extrapolating the luminosities using data from comparable simulations.

In this work, instead, we primarily focus on the results from Ref.~\cite{Fujibayashi:2020dvr}, where neutrino luminosities up to $6~\textrm{s}$ following the merger are provided. Specifically, we adopt the neutrino luminosities from the following three-dimensional merger simulations: DD2-125M, SFHo-125H, and DD2-135M, all from Ref.~\cite{Fujibayashi:2020dvr}. These simulations feature different nuclear equations of state and colliding neutron star masses, and for a representative neutrino luminosity, we take the average over the three. The initial total neutrino luminosity from both the cooling of the neutron star and the surrounding accretion disk is around $10^{53}$ erg/s. While the latter starts falling exponentially around $0.1$ s after the merger, the emission from the newly formed star remains approximately constant throughout $6$ s (see the bottom panel of Figure 3 in \cite{Fujibayashi:2020dvr}). 
It will turn out that electron antineutrinos are most relevant for our analysis. While the authors of Ref.~\cite{Fujibayashi:2020dvr} provide only the all-flavor luminosity, the electron antineutrino luminosity from neutron star mergers is known to dominate over the luminosity from other (anti)neutrino flavors~\cite{Sekiguchi:2011zd,Cusinato:2021zin,Radice:2021jtw,Qiu:2025kgy}.  Taking the results from Ref.~\cite{Fujibayashi:2020dvr} and dividing by $6$ gives the luminosity we adopt for a specific (anti)neutrino flavor, which therefore serves as a conservative estimate for the electron antineutrino luminosity. Since the average neutrino energy is not presented in Ref.~\cite{Fujibayashi:2020dvr}, we use $\langle E_\nu \rangle = 20$~MeV in our calculations, which is roughly the mean of the values considered in Refs.~\cite{Kyutoku:2017wnb,Lin:2019piz}.

\begin{figure}[t!]
    \centering
       \includegraphics[width=\linewidth]{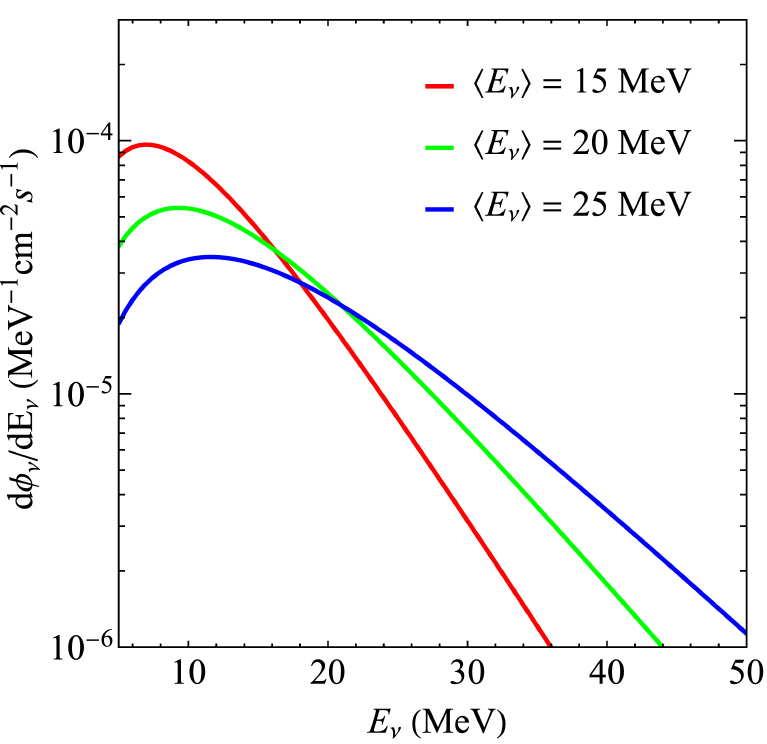}
    \caption{\textcolor{black}{Expected (anti)neutrino flux, for a single flavor, at Earth from binary neutron star mergers up to redshift $z=1$. The flux is shown in the $(5, 50)$ MeV neutrino energy interval for three different average (anti)neutrino energies. We adopt the neutrino luminosities from Ref.~\cite{Fujibayashi:2020dvr}.}}
    \label{fig:flux}
\end{figure}

From the neutrino luminosities discussed above, we construct individual (anti)neutrino fluxes, $\phi_\nu$, using the ``pinching factor'' parametrization from the supernova literature \cite{Keil:2002in,Brdar:2018zds,Chauhan:2022wgj}.
The (anti)neutrino flux at Earth for a particular neutrino flavor from an individual neutron star merger reads
\begin{align}
\left( \frac{d \phi_\nu}{dE_\nu}\right)_\text{single BNS} = &\,\frac{L_\nu}{4\pi d^2} \frac{(1+\alpha)^{1+\alpha}}{\Gamma(1+\alpha)} \frac{E_\nu^\alpha}{\langle E_\nu \rangle^{2+\alpha}}\,\,\nonumber \\ 
&\times \text{exp}\bigg[-(1+\alpha) \frac{E_\nu}{\langle E_\nu \rangle}\bigg] \,.
\label{eq:1merger}
\end{align} 
Here, $L_\nu$ is the (anti)neutrino luminosity of a given flavor, $d$ is the distance from the merger and $\alpha$ is the pinching parameter set to $2.3$, corresponding to the Fermi-Dirac distribution with zero chemical potential.
We checked by varying the value of $\alpha$ in the range of 0 to 5 that the number of predicted neutrino events does not change significantly.
In order to maximize the neutrino flux, we are not focusing on any specific merger; instead, we include the contribution from all the binary neutron star mergers up to redshift $z$. The resulting (anti)neutrino flux at Earth reads
\begin{align}
\frac{d \phi_\nu}{dE_\nu}=\int_0^z (1+z')\, \Phi[E_\nu (1+z')]\, R_{\text{BNS}}(z') \, \bigg|\frac{dt}{dz'}\bigg| \, dz' \,,
\label{eq:allmergers}
\end{align}
where the integration is carried out over redshift due to the cosmological distances.
In this integral, $\Phi$ is defined to be the right-hand side of \cref{eq:1merger} with the $(1/4\pi d^2)$ factor stripped off, and $L_\nu$ integrated over neutrino emission time. Further, $R_{\text{BNS}}(z)$ is the binary neutron star merger rate; we take the ``BNSOpt'' curve from \cite{Lin:2019piz} and rescale it such that $R_{\text{BNS}}(0)$ matches the value $250\,\text{Gpc}^{-3}\,\text{yr}^{-1}$, which is the upper limit on the local neutron star merger rate recently reported by LVK \cite{LIGOScientific:2025pvj}.
In Eq.~\ref{eq:allmergers}, we use
\begin{align}
\left|\frac{dt}{dz}\right|=\Big[H_0 (1+z) \sqrt{\Omega_m \, (1+z)^3+\Omega_\Lambda}\Big]^{-1}\; ,
\end{align}
where $H_0$ is the Hubble constant, and $\Omega_m \approx 0.3$ and $\Omega_\Lambda \approx 0.7$ are the fractions of the total energy density in matter and dark energy, respectively. {\color{black} In~\cref{fig:flux}, we show (anti)neutrino flux at Earth from binary neutron star mergers up to redshift $z=1$, for each individual flavor, and for three different average (anti)neutrino energies}.  Finally, in connection to flux in \cref{eq:allmergers}, let us note that this formula, in the limit $z \to \infty$, also represents the recipe for computing the DSNB flux at Earth, with the binary neutron star merger rate replaced by the core-collapse supernova rate; see e.g. \cite{Ando:2004hc, Beacom:2010kk}. 

 Armed with the (anti)neutrino flux from \cref{eq:allmergers}, we compute the expected number of events in relevant neutrino detectors via
\begin{align}
N_\nu = N T \int_{16~\text{MeV}}^{50~\text{MeV}} \epsilon(E_\nu)\, \sigma(E_\nu)\,\frac{d \phi_\nu}{dE_\nu}\,dE_\nu\,,
\label{eq:event15}
\end{align}
where $T$ is the data-taking time, $N$ is the total number of target nuclei, $\epsilon$ is detection efficiency and $\sigma$ is the neutrino interaction cross section. As a starting point, we consider the JUNO experiment \cite{JUNO:2015zny}, which recently started taking data, as well as the upcoming Hyper-Kamiokande \cite{Hyper-Kamiokande:2018ofw} and DUNE \cite{DUNE:2015lol}. For the former two, the relevant neutrino scattering process in the $\mathcal{O}(10)$ MeV energy range is inverse beta decay $(\bar{\nu}_e + p \to e^+ + n)$, while for DUNE, it is $\nu_e + ^{40}\text{Ar} \to e^- + ^{40}\text{K}^*$. The respective cross sections for these processes are adopted from \cite{Beacom:2010kk,Ricciardi:2022pru,Scholberg:2012id,DUNE:2023rtr}, and 
we also make use of detection efficiencies tabulated in \cite{Tabrizi:2020vmo}. Note that an additional factor of 63.5\% \cite{Li:2024luu}, arising from the fraction of binary neutron star mergers that do not yield a prompt collapse into a black hole, is factored into the efficiency.
Regarding the fiducial mass in the detector, for JUNO we take 17 kt (linear alkylbenzene (LAB) liquid scintillator), 40 kt for DUNE (liquid argon (LAr)), and 374 kt (two tanks) for Hyper-Kamiokande (water).

\begin{figure}[t!]
    \centering
       \includegraphics[width=\linewidth]{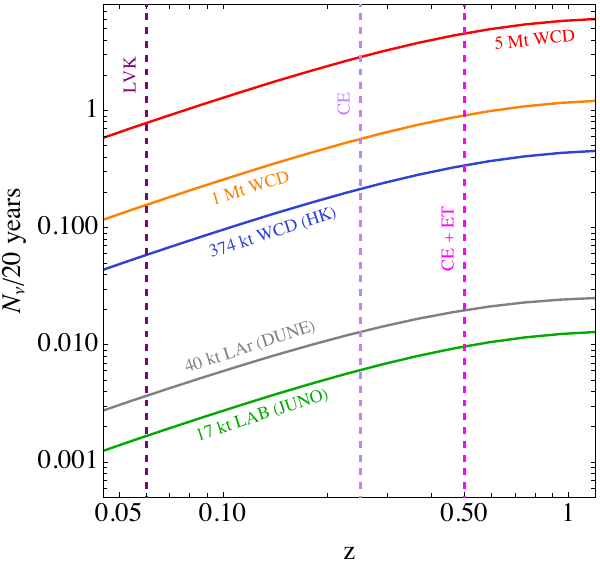}
    \caption{Number of neutrino events from binary neutron star mergers expected in JUNO, DUNE, Hyper-Kamiokande (HK), $1$ Mt and $5$ Mt water Cherenkov detector (WCD) as a function of the maximum redshift considered for such mergers. We adopt the neutrino luminosities from Ref.~\cite{Fujibayashi:2020dvr}. The sensitivities of current and upcoming gravitational-wave interferometers are also shown as vertical lines, indicating the redshift at which a merger can be detected with 90\% efficiency.}
    \label{fig:events}
\end{figure}

In \cref{fig:events} we show the expected number of neutrino events from binary neutron star mergers in the aforementioned detectors over 20 years of data taking as a function of the maximum redshift considered for the sources. We use the total time-integrated luminsoity (up to $6$~s) from Ref.~\cite{Fujibayashi:2020dvr} and take the electron (anti)neutrino average energy to be $20$ MeV. While the cross sections at $\mathcal{O}(10)$ MeV energies are comparable for both inverse beta decay and electron neutrino scattering on argon, the larger number of target nuclei in Hyper-Kamiokande yields a higher event rate compared to DUNE and JUNO. However, even Hyper-Kamiokande falls short of reaching 1 event in $20$ years. Interestingly, with the merger rate of $R_{\text{BNS}}(0)=1700\, \text{Gpc}^{-3}\,\text{yr}^{-1}$ that was still viable after the first three runs of LVK, Hyper-Kamiokande would be expected to make an observation. However, the recent upper limit of $250\, \text{Gpc}^{-3}\,\text{yr}^{-1}$ from the fourth observing run of LVK excludes such a possibility. This leads us towards considering a megaton-scale neutrino detector with an $\mathcal{O}(10)$ MeV threshold energy. Let us note that proposed experiments in this direction include Deep-TITAND \cite{TITANDWorkingGroup:2001qnw,Kistler:2008us}, MEMPHYS \cite{deBellefon:2006vq}, and MICA \cite{Boser:2013oaa}. Motivated by these proposals, we consider water Cherenkov detectors (WCD) of fiducial mass $1$ Mt and $5$ Mt; for detection efficiencies, we adopt the respective values for Hyper-Kamiokande. The results for these two realizations are also shown in \cref{fig:events}, where we observe that a couple of neutrino events from neutron star mergers during a $20$-year period are attainable.


In the figure, we also show vertical lines that indicate the redshift at which present and near-future gravitational-wave detectors can observe binary neutron star mergers with
90\% efficiency \cite{Evans:2021gyd}. We show results for the current LVK network \cite{KAGRA:2021vkt}, the upcoming Cosmic Explorer (CE) \cite{Reitze:2019iox}, and also the combination of Cosmic Explorer and Einstein Telescope (CE+ET) \cite{Abac:2025saz}. Here we assume that Cosmic Explorer will feature a single 20 km observatory, but we note that even more sensitive configurations, including a pair of 40 km observatories, have been discussed \cite{Evans:2021gyd}, and for such a setup the sensitivity to binary neutron star mergers further improves.

\begin{figure}[t!]
    \centering
       \includegraphics[width=\linewidth]{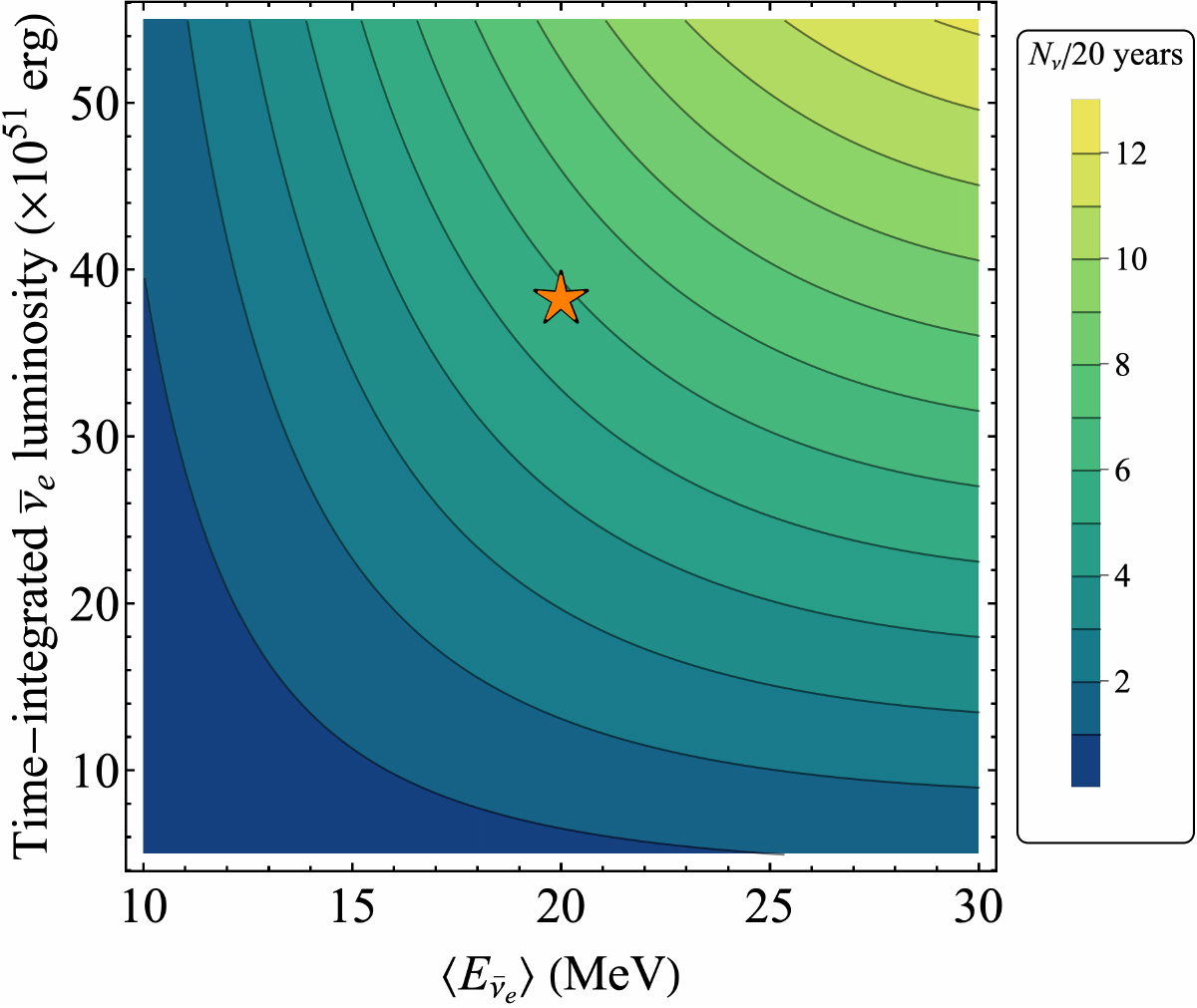}
    \caption{The number of expected neutrino events from binary neutron star mergers up to redshift $z=1$, in the $(16, 50)$ MeV neutrino energy interval at a $5$ Mt WCD over 20 years of data taking. The results are shown as a function of the time-integrated electron antineutrino luminosity and the average energy of electron antineutrinos. The star indicates the values adopted in \cref{fig:events}.}
    \label{fig:lumi}
\end{figure}

Before closing this section, let us stress again that the predicted number of neutrino events scales linearly with the neutrino luminosities. Various simulations in the literature yield neutrino luminosities that differ by up to an order of magnitude. Therefore, we remind the reader that the results shown throughout this work, including \cref{fig:events}, depend explicitly on simulation results associated to Ref.~\cite{Fujibayashi:2020dvr}. There exist simulations in which the predicted luminosities, and consequently the neutrino flux, are noticeably different, and in some cases (see \cite{Lippuner:2017bfm}) even much smaller. To allow the reader to assess, for any given simulation of binary neutron star mergers, whether a sufficient neutrino flux is produced, we show in \cref{fig:lumi} the number of expected neutrino events, with neutrino energies between $16$ and $50$ MeV, at 5 Mt WCD over 20 years of data taking. The results are presented as a function of the time-integrated electron antineutrino luminosity (i.e., total energy released in electron antineutrinos) and the average energy of this species. The orange star in the figure denotes the total time-integrated luminsoity from Ref.~\cite{Fujibayashi:2020dvr} and an average energy of $20$ MeV, used in our analysis.

\begin{figure*}[t] 
    \centering
    \includegraphics[width=\textwidth]{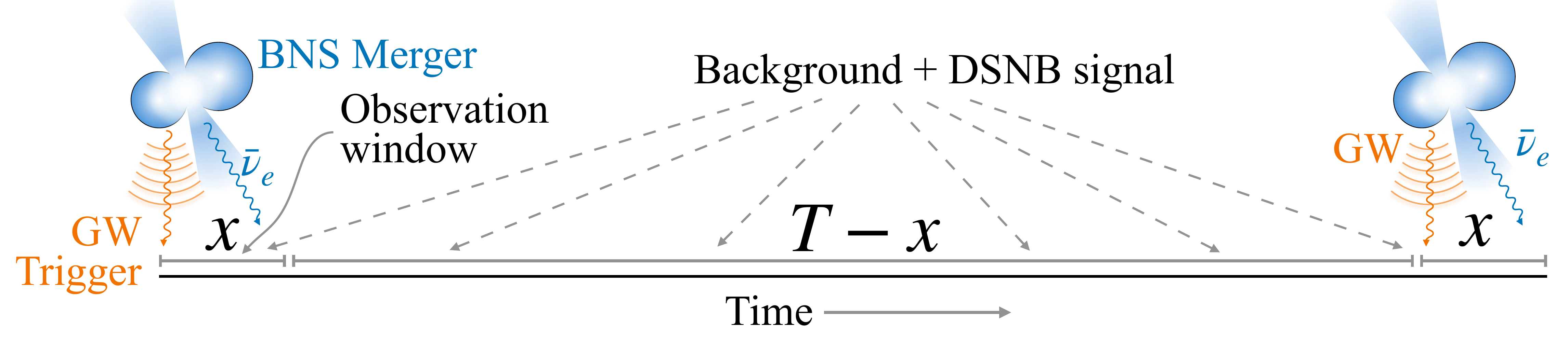}
    \caption{Schematic illustration of the uniform distribution of background events over time, as well as the signal events contained within the observational time window of length~$x$, measured from the coalescence time.}
    \label{fig:schematic}
\end{figure*}

\section{Background mitigation}
\label{sec:bkg}
\noindent
In \cref{fig:events,fig:lumi}, we have presented the expected number of neutrino events from binary neutron star mergers. However, this is by no means the only $\mathcal{O}(10)$ MeV neutrino source, so the detection prospects, in light of different background events, need to be carefully assessed. The backgrounds include, but are not limited to, the DSNB, atmospheric neutrinos, and cosmic ray induced spallation \cite{Super-Kamiokande:2021jaq}. We will present results for the 5 Mt WCD where the total background is taken from Fig.~188 in \cite{Hyper-Kamiokande:2018ofw}, and appropriately rescaled by the fiducial mass ratio between such a detector and Hyper-Kamiokande. The signal detection efficiency is taken to be $\epsilon = 0.9$ \cite{Hyper-Kamiokande:2018ofw,Lin:2019piz}.

According to \cref{fig:events}, a handful of neutrino events from binary neutron star mergers are expected to be detected within $10$ years of observation at a 5~Mt WCD. In the same time period, there will be $\mathcal{O}(10^5)$ background events in such a detector~\cite{Hyper-Kamiokande:2018ofw}. Since all the background events are uncorrelated with gravitational-wave signals, the probability for any particular background event to be recorded within the time period $x$ following any gravitational-wave neutron star coalescence signal is $x/T$; $T$ is the average time interval between two consecutive mergers that yield a long-lived neutron star. This statement is illustrated schematically in \cref{fig:schematic}. In light of the present upper bound on the neutron star merger rate,  $T \simeq 3000~\textrm{s}$ is expected for observation up to a redshift of 0.5, which is accessible at upcoming gravitational wave experiments like Cosmic Explorer~\cite{Evans:2021gyd}.
Since neutrinos from such large distances would be subject to a significant time-of-flight delay due to their nonzero mass, we take $x = 30~\textrm{s}$ as a representative observation window for neutrinos from a merger event with respect to the coalescence time. The probability that none of the $10^5$ background events is misidentified as a neutrino from the merger--or, in other words, the probability that all background events fall outside the time window $x$ for any merger--is $(1 - x/T)^{10^5} = (1 - 30/3000)^{10^5} \approx 3.3 \times 10^{-437}$.

If we now reduce the maximum distance of interest from $z \sim 0.5$ to $z \sim 0.05$, the decrease in the comoving volume as well as the merger rate will result in an overall decrease of the relevant number of mergers by a factor of $1000$. Cosmic Explorer will record such events every $3\times 10^6~\textrm{s}$ (i.e., roughly 1 event per month) instead of every $3000~\textrm{s}$. Further, the decrease in the maximum distance of interest would also lead to a shorter time window of interest, at around 10 s.
The probability of avoiding confusion between background and neutrinos from mergers through their time separation then becomes $\left(1 - 10/(3\times 10^6)\right)^{10^5} \approx 72$\%, which is a striking improvement. The above discussion serves as an illustration of how reducing the maximum considered distance of neutron star mergers can dramatically improve confidence that a neutrino detected shortly after a gravitational-wave signal is indeed coming from neutron star mergers. In what follows, we present the details of our background mitigation analysis.

The time delay of a neutrino emitted at the moment of neutron star coalescence at redshift $z$, relative to the gravitational wave signal of coalescence, is given by \cite{Stodolsky:1999kc}
\begin{align}
\Delta t(z) = \int_0^z \frac{dz'}{H_0 \sqrt{\Omega_m \, (1+z')^3+\Omega_\Lambda}} \frac{m_\nu^2}{2 E_\nu(z')^2}\,, 
\label{eq:timing}
\end{align}
where $m_\nu$ is the mass of the neutrino mass eigenstate under consideration and $E_\nu(z')=(1+z') E_\nu$, with $E_\nu$ being the neutrino energy at Earth. This formula reduces to $\Delta t \approx L \, m_\nu^2/(2 E_\nu^2)$ for galactic-scale distance $L$.

Note that not all the neutrinos from the mergers will be emitted promptly; for instance, the time over which the majority of neutrinos are emitted in simulations performed in Ref.~\cite{Fujibayashi:2020dvr} is about $6~\textrm{s}$. To showcase the effect of the uncertainty in the emission time, we also consider the ``Hinf'' model from Ref.~\cite{Lippuner:2017bfm}, where the emission duration is approximately $0.6$ s, and the total time-integrated luminosity is comparable to the previous scenario~\cite{Lin:2019piz}. { The average emission times in these scenarios are 2.7~s and 0.19~s, respectively.} In our background mitigation analysis, we assume that all neutrinos are emitted at the very end of the emission interval, which adds $6~\textrm{s}$ ($0.6~\textrm{s}$) to the time delay in \cref{eq:timing}, and we denote this quantity as $\Delta t_{\text{tot}}(z)$. The quantity $\Delta t_{\text{tot}}(z)$ is analogous to $x$ in our qualitative picture presented above (see also \cref{fig:schematic}), and maximizing it increases the likelihood of confusing background and signal events, making our analysis conservative.

Regarding the time delay of the neutrino signal relative to the gravitational-wave signal, it is important to calibrate such timing measurements by precisely determining the coalescence time of the merging neutron stars. While LIGO has already demonstrated its capability to detect gravitational waves associated with binary neutron star inspirals \cite{LIGOScientific:2017vwq}, Cosmic Explorer and the Einstein Telescope will significantly improve the precision of binary neutron star coalescence time measurements, reducing the uncertainty to roughly $10~\textrm{ms}$ \cite{Pizzati:2021apa}. The typical neutrino time delay is much larger than this value, implying that, for practical purposes, the coalescence time can be considered precisely measured.\\

We consider neutrino energies in the range of $16$ MeV to $50$~MeV in the detector. For each energy bin of $1$~MeV width, we define the probability $P_{\text{safe}}$ 
for avoiding confusion between neutrinos from the mergers and background events, as
\begin{align}
P_{\text{safe}} = \left( 1-\frac{\Delta t_{\text{tot}}(z_\text{cut})}{T(z_{\text{cut}})} \right)^{n_\text{bkg}(z_\text{cut})}\,.
\label{eq:P_safe}
\end{align}
Here, $n_\text{bkg}$ is the total number of background events in a given energy bin collected in the time interval in which a single neutrino event from mergers, with energy falling in that bin, would be detected. $T(z_\text{cut})$ is the average time between neutron star mergers that do not lead to a prompt black hole collapse, detectable by Cosmic Explorer and the Einstein Telescope, with a maximum merger distance corresponding to a redshift $z_\text{cut}$. The probability $P_{\text{safe}}$ is constructed such that the background-to-signal event ratio in a given bin is $1-P_{\text{safe}}$ for the observation time needed to detect a single neutrino from binary neutron star mergers in that particular energy bin.

Notice that $n_\text{bkg}$ and $\Delta t_{\text{tot}}$ in \cref{eq:P_safe} depend on $z_\text{cut}$. Regarding $\Delta t_{\text{tot}}$, we compute it for a single merger at the average distance of
\begin{align}
r_\text{avg}=\frac{\int_0^{r_{\text{cut}}} dr\, r\, R_\text{BNS} \, 4\pi r^2 }{\int_0^{r_{\text{cut}}} dr\, R_\text{BNS}\, 4\pi r^2} = 0.75 \,r_{\text{cut}} \; ,
\end{align}
considering $R_{\text{BNS}}$ being constant within $\mathcal{O}(1000)~\text{Mpc}$ (lower redshift $z_{\text{cut}} \lesssim 0.2$). The comoving distance depends on the redshift, i.e.,  $r_\text{avg} (z_\text{avg}) = 0.75\, r_\text{cut} (z_\text{cut}) $ implying $ \int_0 ^{z_\text{avg}} \frac{dr}{dz'} dz' = 0.75 \int_0 ^{z_{\text{cut}}} \frac{dr}{dz'} dz'$. The differential comoving distance factor \mbox{$dr/dz=\big[H_0 \sqrt{\Omega_m \, (1+z)^3+\Omega_\Lambda}\big]^{-1}$} remains almost constant at lower redshift, thus yielding $z_\text{avg}=0.75\,z_\text{cut}$. Here, $z_\text{avg}$ is the redshift associated with the average binary neutron star merger distance $r_\text{avg}$, that we use to compute the time delay $\Delta t_\text{tot}$ in \cref{eq:P_safe}, i.e.,
$ \Delta t_\text{tot} (z_{\text{cut}}) = \Delta t(z_\text{avg})\,+$ maximum emission duration.

\begin{figure}[t!]
    \centering
       \includegraphics[width=\linewidth]{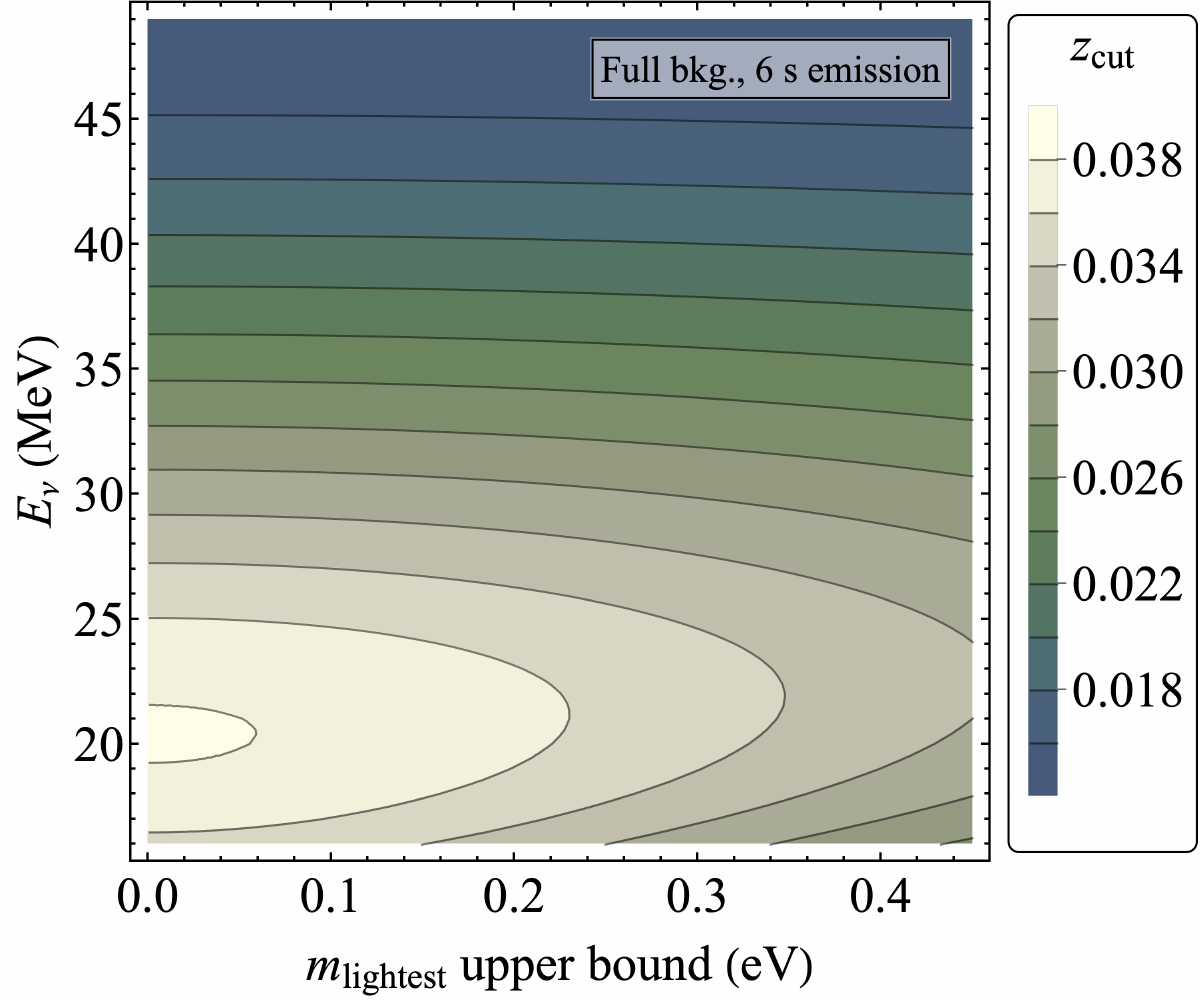}
    \caption{Energy bin-dependent $z_\text{cut}$ as a function of neutrino energy and the lightest neutrino mass for a 5~Mt WCD with the models taken from Ref.~\cite{Fujibayashi:2020dvr}.}
    \label{fig:zcut}
\end{figure}

\begin{figure*}[t!]
\centering

    \includegraphics[width=0.48\linewidth]{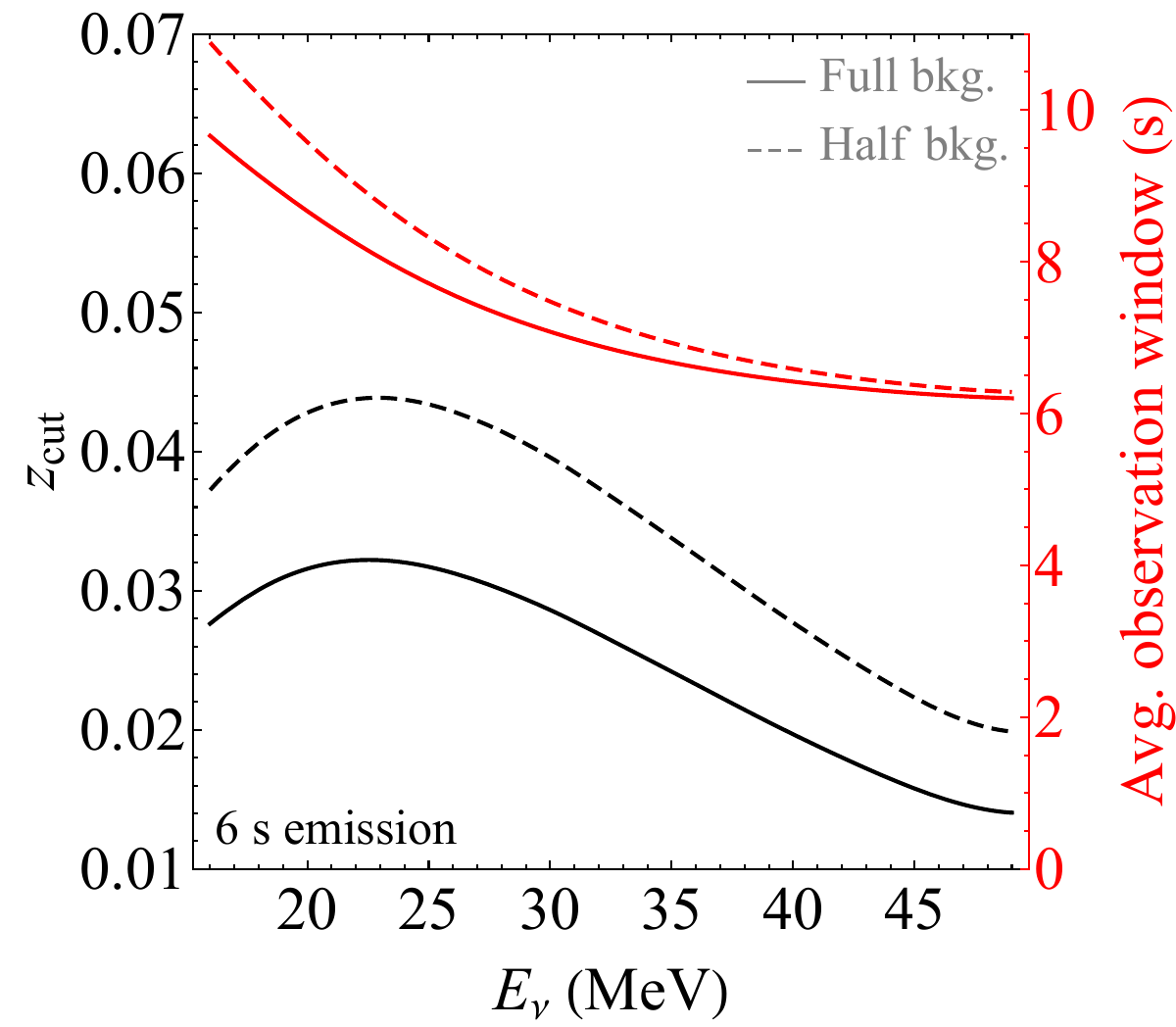}
 \hfill
    \includegraphics[width=0.48\linewidth]{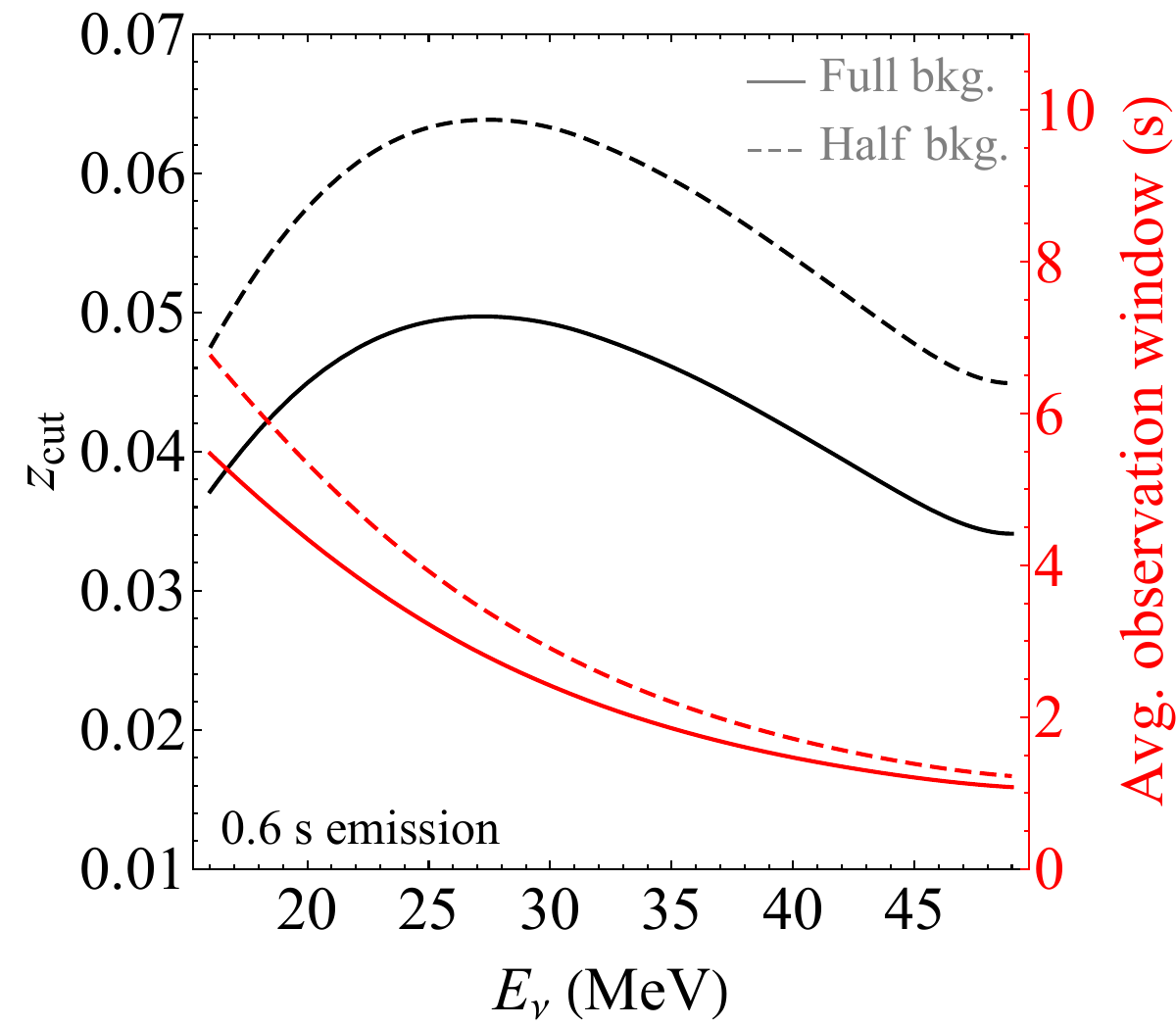}
  \caption{Values of the maximum merger redshift $z_{\text{cut}}$ as a function of neutrino energy for two neutrino emission durations of 6~s (left panel) and 0.6~s (right panel) are shown in black. We also show, in red, the corresponding average value of $\Delta t_{\text{tot}}$. Solid lines show results using the full background expected for a 5~Mt WCD, while dashed lines represent the case of a 50\% background reduction.}
  \label{fig:more_zcut}
\end{figure*}

We determine $z_\text{cut}$ by solving $P_{\text{safe}}=0.9$ for each energy bin, making $z_\text{cut}$ a bin-dependent quantity.
In \cref{fig:zcut}, we show the calculated values of $z_\text{cut}$ for the energy bins $[\, E_\nu,\,E_\nu+1~\text{MeV}\,]$ and the upper bound of the lightest neutrino mass, which enters $\Delta t_{\text{tot}}$ through \cref{eq:timing}; we used the luminosity data and emission time information ($6$ s) corresponding to the simulations in Ref.~\cite{Fujibayashi:2020dvr}. To be conservative, at present, $z_\text{cut}$ values corresponding to the current upper limit of $m_\nu = 0.45$ eV reported by KATRIN should be considered, as this will maximize the neutrino delay. Nevertheless, we illustrate that future improvements in neutrino mass constraints will benefit this analysis, since, as shown in the figure, larger values of $z_\text{cut}$ will become accessible. 

{ Note that, increasing $P_{\text{safe}}$ further, to reach a $2\sigma$ or $3\sigma$ significance would necessitate an even smaller value of $z_\text{cut}$ compared to what is shown in \cref{fig:zcut}. Even for $m_\text{lightest}=0$~eV, value of  $P_{\text{safe}}=0.9545$ ($2\sigma$) requires $z_\text{cut}\approx 0.02$, and $P_{\text{safe}}=0.9973$ ($3\sigma$) requires $z_\text{cut}\approx 0.005$. This would strongly reduce the prospects of observing a neutrino from a BNS merger, since a smaller observational volume would naturally lead to a longer observational time.
}

In \cref{fig:more_zcut}, for the current limit on the lightest neutrino mass at $0.45$ eV, we show $z_\text{cut}$ as a function of neutrino energy both for $6$~s (left panel) and $0.6$~s (right panel) duration of neutrino emission. Solid lines correspond to the case where the background for the 5~Mt WCD is calculated by rescaling the expected background at Hyper-Kamiokande \cite{Hyper-Kamiokande:2018ofw}. We also show, in dashed, the results for the case where the background can be significantly reduced, for instance by doping the detector with gadolinium \cite{Beacom:2003nk}\footnote{{\color{black} The background plays a pivotal role in determination of $z_\text{cut}$, as the number of background events $n_\text{bkg}$ enters into $P_\text{safe}$ in~\cref{eq:P_safe}.
Gadolinium doping can improve signal efficiency significantly by reducing the background. Current status on this detection method can be found in \cite{Koshio:2025fjs}.}}.
As can be seen from the figures, in the latter case, where we assume that 50\% of the background can be removed, our analysis for further background mitigation using gravitational wave and neutrino timing yields larger $z_{\text{cut}}$ values. This is because a smaller initial background allows for $T$ in \cref{eq:P_safe} to be smaller and $\Delta t_{\text{tot}}$ to be larger, compared to the full background case, while still giving $P_\text{safe}=0.9$.
In \cref{fig:more_zcut}, we also show $\Delta t_{\text{tot}}$ from \cref{eq:P_safe}, namely the average duration of the observation window following the merger (corresponding to $z_{\text{avg}} = 0.75 \, z_{\text{cut}}$). We observe, as expected, that the case with longer neutrino emission duration leads to longer $\Delta t_{\text{tot}}$, signifying that in addition to the time delay due to nonzero neutrino mass (\cref{eq:timing}), the emission time window is also playing an important role in determining $z_\text{cut}$. Essentially, the longer emission window ($6$ s) restricts the allowed time-of-flight delay to smaller values, constraining us to smaller $z_{\text{cut}}$ values.

Upon determining $z_\text{cut}$ for each energy bin, we apply \cref{eq:event15} with $z_\text{cut}$ as the upper integration limit in \cref{eq:allmergers}. Finally, we sum the signal events across all bins and, from this, determine the number of years required for the successful detection of a single neutrino from binary neutron star mergers at a 5 Mt WCD with 90\% confidence (i.e., for $P_\text{safe}=0.9$).
Note that allowing $z_\text{cut}$ to vary with energy in our refined search strategy implies that different merger distances should be observed in different energy windows.

\begin{figure}[t!]
    \centering
       \includegraphics[width=0.9\linewidth]{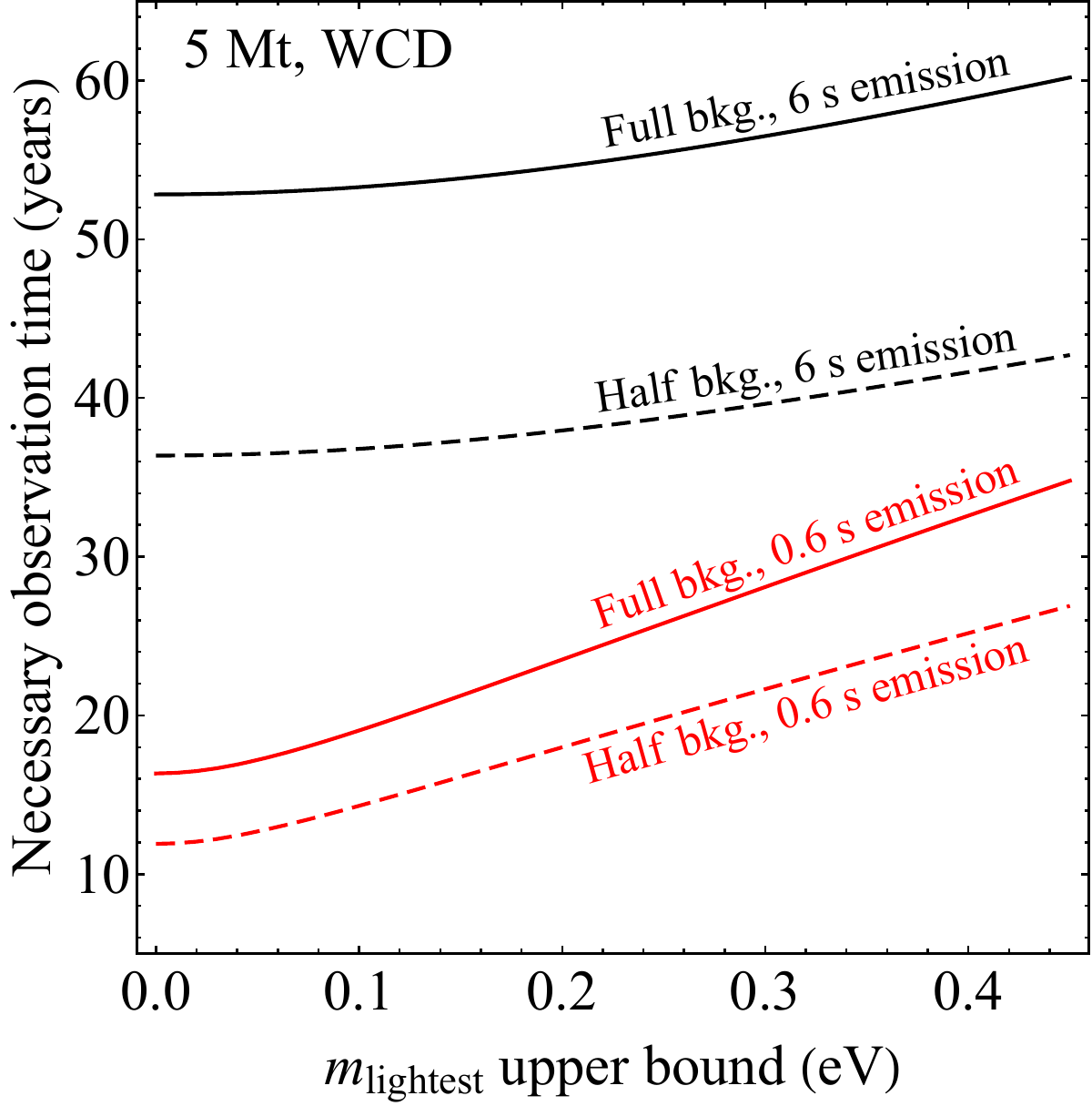}
    \caption{Required observation time for the discovery of a single neutrino from binary neutron star mergers at a 5~Mt WCD as a function of the upper limit on the lightest neutrino mass. Results are shown for emission times of 6~s (black) and 0.6~s (red), for both full background (solid) and a scenario with 50\% background reduction (dashed).}
    \label{fig:time_wait}
\end{figure}

\begin{figure*}[t!]
  \centering
    \includegraphics[height=0.42\textwidth]{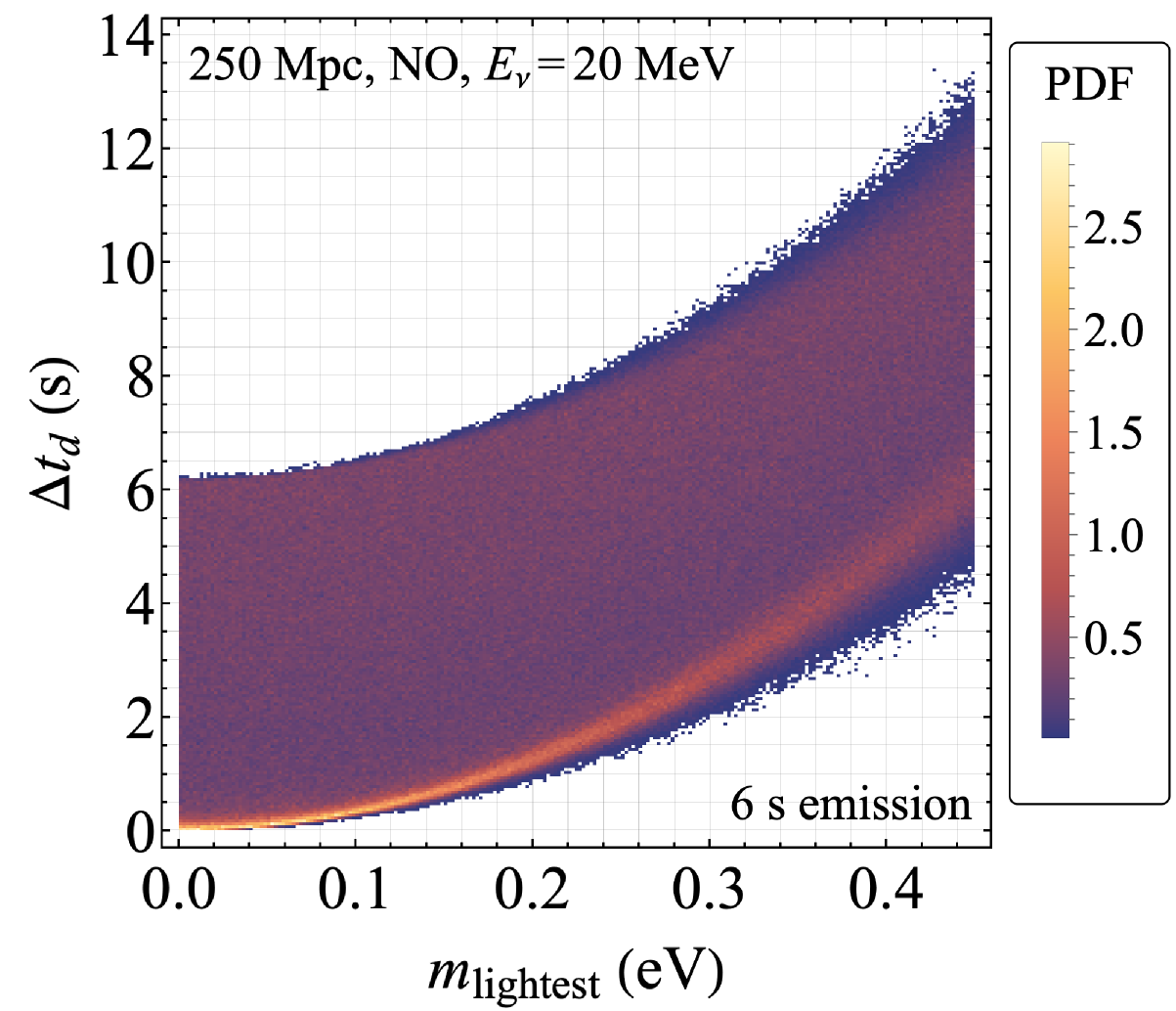}
    \hfill
    \includegraphics[height=0.42\textwidth]{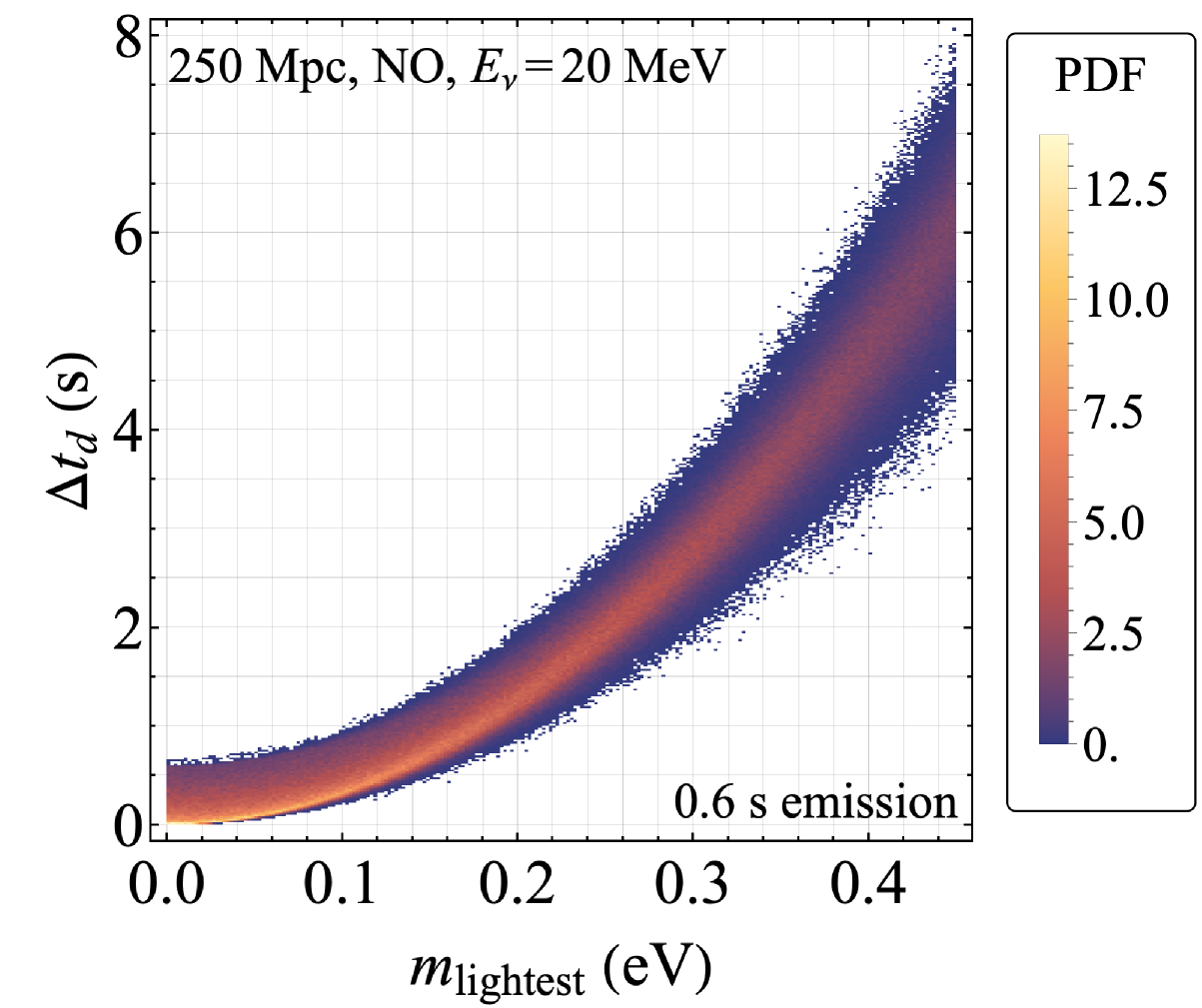}
  \caption{Probability density function in the parameter space of $\Delta t_d$ and the lightest neutrino mass, shown for the two merger simulation models corresponding to neutrino emission durations of $\sim$6~s (left panel) and 0.6~s (right panel).} 
  \label{fig:nu_mass}
\end{figure*}

In \cref{fig:time_wait}, we show the necessary observation time for neutrino discovery from binary neutron star mergers at a 5~Mt WCD with 90\% confidence. We show results for both 6~s (black) and 0.6~s (red) emission times, as well as a scenario with the full background (solid) and the assumption of 50\% background reduction prior to our timing analysis (dashed). The required observation time is shown as a function of the upper limit on the lightest neutrino mass; hence, at present, observation times corresponding to $0.45$~eV apply. For the $0.45$ eV limit, we find that the necessary observation time ranges between $35$ and $60$ years in the vanilla case with the full background; $\sim 30$\% improvements are expected in the case with a two times smaller background. Our findings are more conservative than those in \cite{Lin:2019piz,Deguire:2025zmg}, where the time delay effect caused by nonzero neutrino mass was not considered. Ignoring such effects corresponds to the scenario with a vanishing lightest neutrino mass; in \cref{fig:time_wait}, for the full background scenario, the necessary observation time for $m_\nu=0$~eV is $53$ ($16$) years for the 6~s (0.6~s) emission time.
Therefore, for the shorter emission time, ignoring the neutrino mass effect leads to even $\sim 60$\% shorter required observational times. For longer emission times, the impact of ignoring neutrino mass effects is not very significant, since the time delay is often not competitive with the emission time window and the latter dominates in $\Delta t_\text{tot}$.
Our predictions for the $m_\nu=0$~eV case with full background and 0.6~s emission time agree with Ref.~\cite{Lin:2019piz}. Namely, an exposure of 20~years~$\times$~5~Mt (red solid line, $m_\nu=0$ eV) is comparable to the results from the bottom panels in Fig.~5 of \cite{Lin:2019piz}, computed for a merger rate similar to the current LVK upper limit \cite{LIGOScientific:2025pvj} employed in this work.

In passing, let us also mention that we performed the analysis for the scenario in which we assume that the neutrino mass is precisely measured (which may become a reality in the era of Project 8 \cite{Project8:2022wqh}). For such a case, we find only marginal improvements ($5$\%–$15$\%) on the required observation time with respect to the values reported in \cref{fig:time_wait}.
This arises because a 10\% (at $1\sigma$) uncertainty in the luminosity distance translates, for a fixed nonzero neutrino mass, into a large uncertainty in the predicted arrival time. Further improvements on the luminosity distance uncertainty can mitigate this effect and make the observation time shorter.
Overall, while at this stage the necessary runtime for claiming a neutrino discovery from binary neutron star mergers at $5$ Mt detector is relatively high, being in the ballpark of $50$ years, we stress that future improvements on the neutrino mass limits may lead to a shorter required detector runtime of $20-30$ years.


\section{Neutrino mass sensitivity}
\label{sec:mass}
\noindent
Now that we have established the timescales for successful detection of neutrinos from binary neutron star mergers, we can ask what kind of physics implications such a discovery could bring. Clearly, with a single neutrino from such sources, there is no cross-section or flavor transition analysis to be done, as those require high statistics. In contrast, a single neutrino event is sufficient, in principle, for studying a time delay of neutrinos with respect to the gravitational-wave signal. This is precisely what we will discuss in what follows by showing how well the neutrino mass can be constrained using time-of-flight effects.

Before proceeding, let us stress the difference between the time delay effect discussed in \cref{sec:bkg} and what we present in this section. There, we have shown that time delay effects control $\Delta t_\text{tot}$, which plays an important role in background mitigation. In that analysis, we adopted the present bound on the neutrino mass, which conservatively maximizes $\Delta t_\text{tot}$ in which we expect neutrinos to arrive relative to the gravitational-wave coalescence signal.
Therefore, the observed time delay will always be smaller than $\Delta t_\text{tot}$.
Given the identification of a single neutrino from binary neutron star mergers through such analysis, we can, in the next step, employ the time delay of this neutrino event to calculate the constraint on the neutrino mass. In what follows, we derive the expected sensitivity.

Based on our $z_\text{cut}$ analysis, we do not expect to record a neutrino from a binary neutron star merger at $z\gtrsim 0.07$. Therefore, for the merger yielding the observed neutrino, detectable in a 5~Mt detector, we take a representative distance of 250~Mpc.
We take the neutrino energy to be $20$~MeV. Since the time delay, introduced in \cref{eq:timing}, at the considered distances reads approximately $\Delta t \approx L \,m_\nu^2/(2 E_\nu^2)$, our results can be mapped to any combination of energy and distance, given the $L/E_\nu^2$ dependence.

We perform a uniform scan over the lightest neutrino mass in the range $(0,\,0.45)$ eV, and for each mass, we generate neutrino time delay values, denoted as $\Delta t_d$; overall, we generate $10^6$ values.
This time interval is the sum of the delay due to the nonzero neutrino mass (see \cref{eq:timing}) and the emission time relative to coalescence; the latter is drawn from time-dependent luminosity profiles in Ref.~\cite{Fujibayashi:2020dvr} ($\sim$6~s total emission) and Ref.~\cite{Lippuner:2017bfm} (0.6~s). Note the distinction between $\Delta t_d$ and $\Delta t_\text{tot}$ employed in \cref{sec:bkg}: while both include the time delay due to neutrino mass, in $\Delta t_\text{tot}$ we included the full duration of the emission time, conservatively maximizing the neutrino time delay. We also account for the uncertainty in the measured distance from binary neutron star mergers.
For each of the sampled points, the distance is drawn from a Gaussian distribution, with a mean value of $250$ Mpc, and $\sigma = 25$ Mpc, which is 10\% of the mean distance.
Note that the distance to GW170817 was measured with a $\sim$25\% uncertainty \cite{LIGOScientific:2017vwq}, {\color{black} further precision was later achieved by applying the surface brightness fluctuation technique to the host galaxy NGC4993~\cite{Cantiello:2018ffy}. Such an approach may also help reduce the distance uncertainty for future BNS mergers.} Significant improvements are expected in the era of Cosmic Explorer and Einstein Telescope.
At a 5~Mt WCD, the relevant neutrino scattering process is inverse beta decay, which involves electron antineutrinos.
Transforming to the mass basis and weighting with $|U_{ei}|^2$ (PMNS matrix elements squared), we calculate the relative numbers of $\nu_1$, $\nu_2$, and $\nu_3$ interactions in our sample.
In normal neutrino mass ordering, $\nu_1$ corresponds to the lightest neutrino mass that we scan over, while the heavier two are computed using the precisely measured neutrino mass-squared differences from oscillation experiments. We present results for normal ordering only, with calculations for the inverted ordering differing only marginally.

\begin{figure}[t!]
    \centering
       \includegraphics[width=\linewidth]{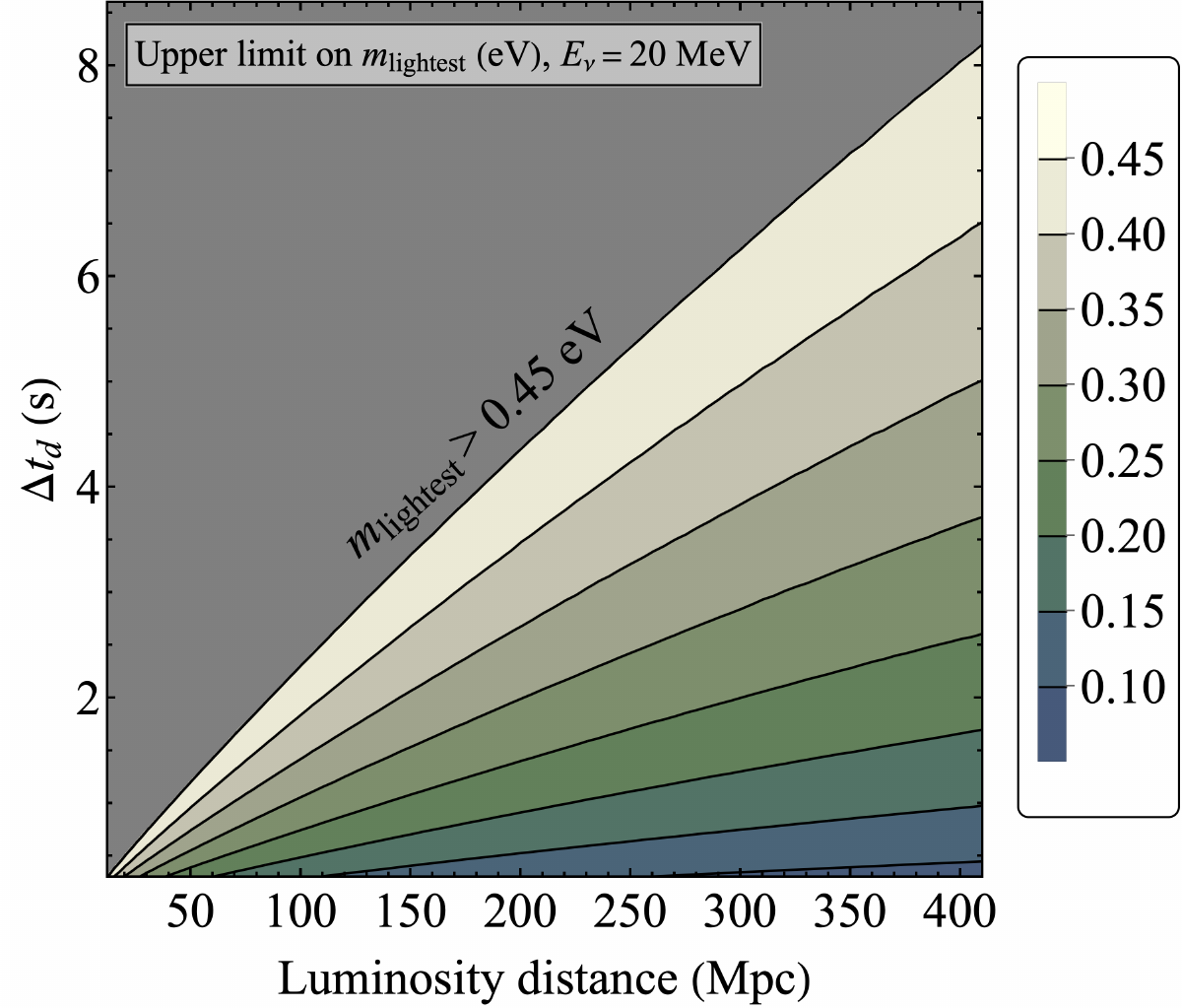}
    \caption{Expected 90\% CL upper limit on the lightest neutrino mass, shown for various combinations of $\Delta t_d$ and merger distance, for the $6$~s emission duration profile.}
    \label{fig:upper_limit}
\end{figure}

In \cref{fig:nu_mass}, we show the results of the analysis described above for the two neutrino emission models with durations of 6~s (left panel) and 0.6~s (right panel). The panels display the probability density function in the parameter space of $\Delta t_d$ and the lightest neutrino mass. We observe that the spread in $\Delta t_d$ is much larger in the 6~s emission case, illustrating that a detected neutrino could have been produced at any time within the 6~s emission interval. Observations from gravitational wave and neutrino experiments will provide $\Delta t_d$ for a neutrino detected from a binary neutron star merger.
From this, one can infer the possible range of the lightest neutrino mass: for a horizontal line indicating a given value of $\Delta t_d$, the two intersections of the line with the edges of the probability distribution allow the neutrino mass limits to be extracted.
Note that with gravitational wave and neutrino timing, both lower and upper limits can be obtained. 
The sensitivity for extracting the lower limit on the mass of the lightest neutrino is a unique feature arising from binary neutron star mergers.
However, the lower limit is highly dependent on the emission time, as illustrated by comparing the panels in \cref{fig:nu_mass}, implying that its determination requires knowledge of the neutrino emission profile.

Therefore, in what follows, we focus on the expected 90\% CL upper limit on the lightest neutrino mass, as presented in \cref{fig:upper_limit}, where we show results for various combinations of $\Delta t_d$ and merger distance for a neutrino of 20 MeV energy, in normal mass ordering (for the $6$~s neutrino emission duration profile).
As mentioned previously, for an observation of a binary neutron star neutrino with a different energy, the relevant $\Delta t_d$ will be approximately scaled as $(20~\text{MeV}/E_\nu)^2$.
As illustrated, there is a wide parameter space where the resulting neutrino mass limit, following the observation of a neutrino from a neutron star merger, would significantly improve upon the present constraint from KATRIN. It is expected that we will gain a better understanding and agreement between cosmological mass limits, and the bound from KATRIN is also expected to improve with the Project~8 experiment.
However, if improving the current limit from terrestrial experiments, supernova, and cosmological data proves to be challenging, a 5~Mt detector could detect neutrinos from binary neutron star mergers and provide leading constraints on the neutrino mass.\\

\section{Conclusions}
\label{sec:conclusions}
\noindent
In this work, we first investigated the prospects for discovering neutrinos from binary neutron star mergers. We found that the expected event rates at current and near-future neutrino experiments will likely not yield a detection, and that detectors at the Mt-scale are necessary. While this may appear to be a distant goal, note that the fiducial mass of upcoming Hyper-Kamiokande exceeds that of Kamiokande-II, which detected the SN1987A neutrinos, by two orders of magnitude. The standard for neutrino detector mass has therefore been increasing rapidly, and Mt-scale detectors based on existing water-Cherenkov technology appear feasible.
We performed a detailed background mitigation analysis, using the coincidence between the gravitational wave and neutrino signals from the mergers, and found that at least several decades of operation are required for discovery, for a 5~Mt neutrino detector.
The search strategy essentially boils down to focusing on the closest and the most neutrino-luminous $\mathcal{O} (10^2-10^3)$ binary neutron star mergers.

Our predictions are more conservative than earlier studies, because the updated LVK O4a results impose a stricter bound on the binary neutron star merger rate, and because earlier analyses did not account for neutrino-mass–induced time delays or for potentially longer emission durations.
Finally, we showed that, following a successful detection, neutrino events from binary neutron star mergers can also be used to constrain the neutrino mass. We find that the expected upper limits can significantly improve upon the present bounds, and binary neutron star mergers may become essential for probing the neutrino mass if terrestrial and supernova probes of neutrino mass do not provide substantial improvement in the future.

\section*{Acknowledgments}
\noindent
The work of VB and SRM is supported by the United States Department of Energy under Grant No.~DE-SC0025477. We acknowledge support from the National Science Foundation 
(Grant No.~NSF-2349581), which enabled MR to conduct research at OSU through the REU program.

\bibliographystyle{JHEP}
\bibliography{refs}

\end{document}